\documentclass{article}
\usepackage[vcentermath]{youngtab}
\usepackage{alltt}
\usepackage[all]{xy}
\usepackage{latexsym,amssymb,amsmath,amsfonts, amsthm, xcolor}

\def\vbar{\mathchoice{\vrule height6.3ptdepth-.5ptwidth.8pt\kern-.8pt}
  {\vrule height6.3ptdepth-.5ptwidth.8pt\kern-.8pt}
  {\vrule height4.1ptdepth-.35ptwidth.6pt\kern-.6pt}
  {\vrule height3.1ptdepth-.25ptwidth.5pt\kern-.5pt}}
\def\fudge{\mathchoice{}{}{\mkern.5mu}{\mkern.8mu}}
\def\bbc#1#2{{\rm \mkern#2mu\vbar\mkern-#2mu#1}}
\def\bbb#1{{\rm I\mkern-3.5mu #1}}
\def\bba#1#2{{\rm #1\mkern-#2mu\fudge #1}}
\def\bb#1{{\count4=`#1 \advance\count4by-64 \ifcase\count4\or\bba A{11.5}\or
  \bbb B\or\bbc C{5}\or\bbb D\or\bbb E\or\bbb F \or\bbc G{5}\or\bbb H\or
  \bbb I\or\bbc J{3}\or\bbb K\or\bbb L \or\bbb M\or\bbb N\or\bbc O{5} \or
  \bbb P\or\bbc Q{5}\or\bbb R\or\bbc S{4.2}\or\bba T{10.5}\or\bbc U{5}\or
  \bba V{12}\or\bba W{16.5}\or\bba X{11}\or\bba Y{11.7}\or\bba Z{7.5}\fi}}

\newcommand{\vs}{\vspace{0.25cm}}

\newtheorem{theorem}{Theorem}
\newtheorem{itlemma}{Lemma}[section]
\newtheorem{itproposition}[itlemma]{Proposition}
\newtheorem{itcorollary}[itlemma]{Corollary}
\newtheorem{itremark}[itlemma]{Remark}
\newtheorem{itremarks}[itlemma]{Remarks}
\newtheorem{itdefinition}[itlemma]{Definition}
\newtheorem{itexample}[itlemma]{Example}

\newenvironment{lemma}{\begin{itlemma}\rm}{\end{itlemma}} %no-italics
\newenvironment{remark}{\begin{itremark}\rm}{\end{itremark}} %no-italics
\newenvironment{remarks}{\begin{itremarks} \rm}{\end{itremarks}}
\newenvironment{corollary}{\begin{itcorollary}\rm}{\end{itcorollary}}
\newenvironment{proposition}{\begin{itproposition}\rm}{\end{itproposition}}
\newenvironment{definition}{\begin{itdefinition}\rm}{\end{itdefinition}}
\newenvironment{example}{\begin{itexample}\rm}{\end{itexample}}
\newenvironment{fact}{\noindent {{\bf Fact}}:\ \ }{\hfill \medskip}
\newenvironment{claim}{\noindent {\em Claim}. \ \ }{\hfill \medskip}
\newcommand{\be}[1]{\begin{equation}\label{#1}}
\newcommand{\ee}{\end{equation}}
\newcommand{\bl}[1]{\begin{lemma}\label{#1}}
\newcommand{\br}[1]{\begin{remark}\label{#1}}
\newcommand{\brs}[1]{\begin{remarks}\label{#1}}
\newcommand{\bt}[1]{\begin{theorem}\label{#1}}
\newcommand{\bd}[1]{\begin{definition}\label{#1}}
\newcommand{\bp}[1]{\begin{proposition}\label{#1}}
\newcommand{\bc}[1]{\begin{corollary}\label{#1}}
\newcommand{\bfact}[1]{\begin{fact}\label{#1}}
\newcommand{\bex}[1]{\begin{example}\label{#1}}
\newcommand{\ec}{\end{corollary}}
\newcommand{\efact}{\end{fact}}
\newcommand{\eex}{\end{example}}
\newcommand{\el}{\end{lemma}}
\newcommand{\er}{\end{remark}}
\newcommand{\ers}{\end{remarks}}
\newcommand{\et}{\end{theorem}}
\newcommand{\ed}{\end{definition}}
\newcommand{\ep}{\end{proposition}}
\newcommand{\epr}{\end{proof}}
\newcommand{\bpr}{\begin{proof}}
\newcommand{\bcl}{\begin{claim}}
\newcommand{\ecl}{\end{claim}}

\newcommand{\bi}{\begin{itemize}}
\newcommand{\ei}{\end{itemize}}
\newcommand{\ben}{\begin{enumerate}}
\newcommand{\een}{\end{enumerate}}
\usepackage{graphicx}

\title{\bf \Large{SUBSPACE CONTROLLABILITY OF BIPARTITE SYMMETRIC  SPIN NETWORKS UNDER GLOBAL CONTROL}
}

\author{Francesca Albertini$^{1}$  \,  and \, Domenico D'Alessandro$^{2}$
\thanks{$^{1}$ Francesca Albertini is with Dipartimento di Tecnica e Gestione 
dei Sistemi Industriali,  Universit\`a di Padova,  {\tt\small francesca.albertini@unipd.it} }
\thanks{$^{2}$ Domenico D'Alessandro is with Department of Mathematics, Iowa State University, Ames, Iowa, U.S.A., {\tt\small daless@iastate.edu}}
}
\begin{document}

\maketitle
%\thispagestyle{empty}
%\pagestyle{empty}

%%%%%%%%%%%%%%%%%%%%%%%%%%%%%%%%%%%%%%%%%%%%%%%%%%%%%%%%%%%%%%%%%%%%%%%%%%%%%%%%
\begin{abstract}

We consider a class of spin networks where each spin in a certain set interacts, via Ising coupling, with a set of {\it central} spins, and the control acts simultaneously on all the spins. This is a common situation for instance in NV centers in diamonds, and we focus on the 
physical case of up to two central spins.  Due to the permutation symmetries of the network, the system is not globally controllable 
but it displays invariant subspaces of the underlying Hilbert space. The system is said to be {\it subspace controllable} if it is controllable on each of these subspaces. We characterize the given invariant subspaces and  the dynamical Lie algebra of this class of systems and prove subspace controllability in every case. 

\end{abstract}

%%%%%%%%%%%%%%%%%%%%%%%%%%%%%%%%%%%%%%%%%%%%%%%%%%%%%%%%%%%%%%%%%%%%%%%%%%%%%%%%
\vspace{0.5cm}


\vspace{0.5cm}

\section{Introduction}
Controllability of finite dimensional quantum systems, described by a 
Schr\"odinger equation of the form 
\be{QSbasic}
|\dot \psi \rangle =(A+\sum_{j} B_j u_j(t)) |\psi\rangle, 
\ee
is usually assessed by computing the Lie algebra ${\cal G}$ generated by the matrices in $u(N)$, $A$ and $B_j$ (see, e.g., \cite{Mikobook}, \cite{JS}, \cite{Salapaka}). The Lie algebra ${\cal G}$  is called the {\it dynamical Lie algebra }. Here $u_j=u_j(t)$ are the (semiclassical) control electromagnetic fields and $|\psi\rangle$ is the quantum mechanical state varying in a Hilbert space ${\cal H}$. If $e^{\cal G}$ denotes  the connected component containing the identity  of the Lie group associated with ${\cal G}$,   then the set of states reachable from $|\psi_0 \rangle$ by choosing the control fields is (dense in) $\{ |\psi \rangle:= X |\psi_0\rangle \in {\cal H} \, | \, X \in e^{\cal G} \}$. In particular if ${\cal G}=u(N)$ or ${\cal G}=su(N)$, the system is said to be {\it (completely) controllable} and every unitary operation, or special unitary operation in the $su(n)$ case, can be performed on the quantum state. This is important in quantum information processing \cite{NC} when we want to ensure that every quantum operation  can be obtained for a certain physical experiment (universal quantum computation). 
Although controllability is a generic property (see, e.g., \cite{Lloyd}), often symmetries of the physical system prevent it and the dynamical 
Lie algebra ${\cal G}$ is a {\it proper} subalgebra of $su(N)$. In this case,  the given representation of the Lie algebra ${\cal G}$ splits into its irreducible components which all act on an invariant subspace of the full Hilbert space ${\cal H}$ on which the system state $|\psi \rangle$ is defined. It is therefore of interest to study whether, on each subspace, controllability is verified, so that, in particular, one can perform universal quantum computation and-or generate interesting  states on a smaller portion of the Hilbert space (see, e.g.,  \cite{W}, \cite{GHZ}). This situation has  recently been studied in detail for networks of particles with spin in the papers \cite{ConDaniel}, \cite{SenzaDaniel}. In particular,  in \cite{SenzaDaniel}, various topologies of  the spin network were considered for various possible interactions among the spins and results were proven concerning the controllability of the {\it first excitation space}, that is, the invariant subspace of the network of states of the form $\sum_ja_j|000\cdots00100\cdots000\rangle$, i.e., superpositions of states where only one spin is in the excited state. In \cite{ConDaniel},  only {\it chains} with next neighbor interactions were considered (instead of general networks)  but 
comprehensive controllability results were given on {\it all} the invariant  subspaces of this type of systems. In both these papers,  the control affects {\it only one} of the spins in the network, which may be placed  in various places in the network. 
%One interesting achievement of the paper \cite{ConDaniel} is to show that the dimension of one of the invariant subspaces of the system still grows {\it exponentially} with the number of spins. Given that subspace controllability is proven on this subspace, this opens the door to achieving universal quantum computation with the given set-up, in particular with physical control on a single spin. 

\begin{figure}[htb]
\centering
\includegraphics[width=0.65\textwidth, height=0.35\textheight]{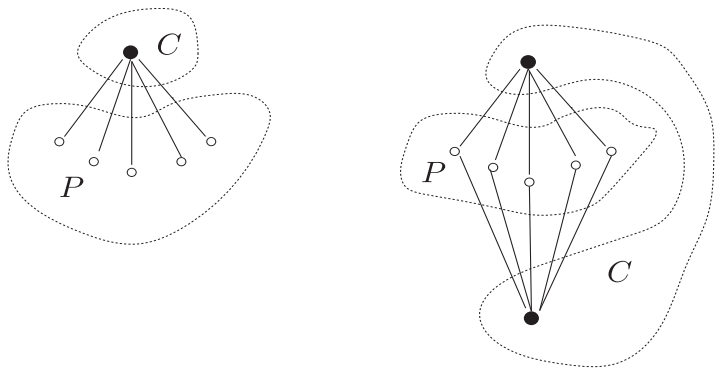}
\caption{Schematic representation of a spin network with one (a) and two (b) central spins $C$ depicted with black bullets as opposed to empty circles (spins in $P$)}.
\label{Fig1}
\end{figure}

The present paper is motivated by experimental situations where control on a single spin particle is not possible and all the  spins of the network are  controlled 
{\it simultaneously}. We want to study the structure of the dynamical Lie algebra and subspace controllability in this situation. We shall consider the case where the spins of the network are arranged in two sets, a set $P$ and a set $C$. The set $C$ is called of {\it central} spins. Spins in the set $P$ ($C$) interact in the same 
(Ising) way with the set of { spins} in the set  $C$ ($P$) but do not interact with each other. The systems we have in mind are, for instance, $N-V$ center in diamonds  \cite{Slava1} \cite{Slava2}, where one or two  central spins (of type $C$),  interact in the same way (via Ising interaction) with a 
bath  of surrounding  spins as in Figure \ref{Fig1}. 
From a mathematical standpoint, such a situation 
can be  extended to the case where there is an arbitrary number of spins in the sets $C$ and $P$. However the interaction between spins is physically a function of the type of spins and of the {\it distance} between 
the spins. It is therefore impossible to have three or more spins in both sets $C$ and $P$ and therefore  we assume that the set with smaller  cardinality,  which we assume to be $C$, has at most two spins. Systems of this type admit {\it symmetries}. In particular, by permuting the spins in the set $C$ and-or the spins in the set  $P$, the Hamiltonian describing the dynamics of the system as  in (\ref{QSbasic}) is  left unchanged (see next section for details). Then, if $n_c$ is the cardinality of the set $C$ and $n_p$ is the cardinality of the set $P$, the group of symmetries is the product between the symmetry group on $n_c$ elements, $S_{n_c}$, and the symmetry group on $n_p$ elements, $S_{n_p}$. In this context, the results of this paper  are the first step towards developing a theory for controllability of spin networks where symmetries are `localized' within certain subsets of the network.

In general terms, if there is a discrete group $G$ of  symmetries  for a quantum mechanical system, the dynamical Lie algebra ${\cal G}$ associated  with the system will be a subalgebra of ${\cal L}^G$, the largest subalgebra of $u(N)$ (N being the dimension of the system) which commutes with $G$. If ${\cal G}$ is {\it equal} to ${\cal L}^G$, subspace controllability is satisfied for each of the invariant subspaces of the 
system (cf. Theorem 2 in \cite{Jonas}). However ${\cal G}$ might be a {\it proper} Lie subalgebra  of ${\cal L}^G$ and subspace 
controllability may  not be satisfied. For the systems we consider in this paper we will see  that ${\cal G}$ is not exactly equal to ${\cal L}^G$. However, this does not affect the subspace controllability of the system for each of its invariant subspaces which, we will prove,  is still verified.  

The controllability of spin networks where one can permute the spins arbitrarily ({\it completely symmetric spin networks}) was studied in \cite{NoiGHZ} expanding upon a study that was started in \cite{Xinhua} motivated by \cite{W}, \cite{GHZ}. In \cite{Jonas}, it was shown how to use {\it Generalized Young Symmetrizers} for the group $G$ to characterize ${\cal L}^G$ in every case, extending some of the results of \cite{NoiGHZ}  to higher dimensions. We shall use the results of these works in 
the following. 

The paper is organized as follows. In the next section, we set up the notations and the basic definitions, so that we can precisely describe  the model we want to treat and the problem we want to consider. We 
also prove a number of preliminary results which will be used later in the paper. The main results are given in section \ref{DDLA} where we describe the dynamical Lie algebra for Ising 
networks of spins with one or two  central spin under global control. Subspace controllability will come as a consequence of this in section \ref{SC3}. Some concluding remarks on the given results will be given in section \ref{conclu}.

\section{Preliminaries}

\subsection{Notations, Basic Definitions and Properties}\label{NBD}
In the following,  we will have to compute a basis for a Lie algebra generated by a given set of matrices. In these calculations, it is not important if we obtain a matrix $A$ or a matrix $kA$ with $k \not=0$. Therefore we shall use the notation $[A,B]\vDash D$ to indicate that the commutator of $A$ and $B$ ($[A,B]:=AB-BA$) is $kD$ for some $k \not= 0$ and therefore $D$ belongs to the Lie algebra that contains $A$ and $B$. We shall also 
often use the formula 
$$
[A \otimes B, C \otimes D]=\frac{1}{2}\{A,B\} \otimes [B,D]+\frac{1}{2} [A,C] \otimes 
\{B,D\}, 
$$
where $\{A,B\}$ denotes the {\it anti-commutator} of $A$ and $B$, i.e.,   
$\{A,B\}:=AB+BA$.  We will do this routinely without explicitly referring to this formula. In $u(n)$ we shall use the {\it inner product} $\langle A, B \rangle:=Tr(AB^\dagger)$. One property of this inner product which will be useful is given by 
the following: 

\bl{commut56} If $A$  commutes with $B$ and $C$, then it is also orthogonal and commutes with $[B,C]$. 
\el   
\bpr Commutativity follows from the Jacobi identity.  Moreover, $Tr(A[B,C]^\dagger)=-Tr(A[B,C])=-Tr(ABC-ACB)=-Tr(BAC-CAB)=-Tr(BAC-BCA)=-
Tr(B[A,C])=0$. 
\epr

The {\it Pauli matrices} $\sigma_{(x,y,z)}$ are defined as 
\be{PauliMat}
\sigma_x:=\begin{pmatrix} 0 & 1 \cr 1 & 0  \end{pmatrix}, \,  \sigma_y:=
 \begin{pmatrix} 0 & i \cr -i & 0  \end{pmatrix}, \, \sigma_z:= \begin{pmatrix}  1 & 0 \cr 0 & -1\end{pmatrix}.
\ee
If  $\bf{1}$ denotes  the identity matrix, 
the Pauli matrices  satisfy 
\be{relations}
\begin{array}{c}
\sigma_x\sigma_x=\sigma_y\sigma_y=\sigma_z\sigma_z={\bf 1}_2, \\
\begin{array}{ccc}
\sigma_x\sigma_y=-i \sigma_z,&  \sigma_y\sigma_z=-i \sigma_x, & \sigma_z\sigma_x=-i \sigma_y \\
\sigma_y\sigma_x=i \sigma_z,& \sigma_z\sigma_y=i \sigma_x, &  \sigma_x\sigma_z=i \sigma_y \\
\end{array} \\
\end{array},
\ee
which give the commutation relations 
\be{commurel}
[i\sigma_x, i \sigma_y]=2i\sigma_z, \, [i\sigma_y, i \sigma_z]=2i\sigma_x, 
\,  [i\sigma_z, i \sigma_x]=2i\sigma_y.
\ee
We shall use  ${\bf 1}$ for the identity matrix in different dimensions as the dimensions will be clear from the context. 
In the most general setting, our model consists of $n=n_c+n_p$ spin $\frac{1}{2}$ particles, with $n_c$ of a type $C$ (for example $n_c$ nuclei) and $n_p$ of the type $P$ (for example $n_p$ electrons). In our conventions, the first $n_c$ positions in a 
tensor product refer to operators on the spins in the set $C$, while the 
following $n_p$ refer to operators on the set $P$. Our main results on the characterization of the dynamical Lie algebra and subspace controllability will concern the physical case of $n_c=1$ and $n_c=2$ and we shall assume without loss of generality $n_c \leq n_p$. We  start giving some general results valid for arbitrary $n_c$. 

We denote by $S_{(x,y,z)}^{C(P)}$ the sum of  $n_{c(p)}$ tensor products $\sum_{j=1}^{n_{c(p)}} {\bf 1}\otimes \cdots \otimes \sigma_{(x,y,z)} \otimes {\bf 1} \otimes \cdots \otimes {\bf 1}$ where the Pauli matrix $\sigma_{(x,y,z)}$ varies 
among all the possible $n_{c(p)}$ positions. For example, if $n_c=2$, $S^C_x:=\sigma_x \otimes {\bf 1}+ {\bf 1} \otimes \sigma_x$. When it is not important or it is clear  whether we refer to the set $C$ or the set $P$, we shall simply denote this  
type of matrices by $S_{(x,y,z)}$. In particular matrices on the left (right)  of a tensor product always refer to operators on the set $C$ ($P$). We notice that $S_{(x,y,z)}$ satisfy 
the same commutation relations as $\sigma_{(x,y,z)}$ and therefore $iS_{(x,y,z)}$ give a representation of $su(2)$ in the appropriate dimensions. We shall denote the $3$-dimensional Lie algebra spanned by $iS_{(x,y,z)}$ with ${\cal S}$. We shall also denote by $I_{(x,y,z)(x,y,z)}^{C(P)}$ matrices which are sum of the tensor products of $2 \times 2$ identities, ${\bf 1}$, except in all possible pairs of 
 positions which are occupied by $\sigma_{(x,y,z)}$ and  $\sigma_{(x,y,z)}$. For example, if $n_c=3$, we have
 $$
I_{xx}^C:=\sigma_x \otimes \sigma_x \otimes {\bf 1} + \sigma_x \otimes {\bf 1} \otimes \sigma_x + {\bf 1}\otimes \sigma_x \otimes \sigma_x, $$ $$ I_{xy}^C:=\sigma_x \otimes \sigma_y \otimes {\bf 1} + \sigma_y \otimes \sigma_x \otimes {\bf 1}+
\sigma_x \otimes {\bf 1} \otimes \sigma_y +
$$
$$+ \sigma_y \otimes {\bf 1} \otimes \sigma_x+ 
{\bf 1} \otimes  \sigma_x \otimes \sigma_y + {\bf 1} \otimes  \sigma_y 
\otimes \sigma_x. 
$$
As before, when it is not important, or it is clear  in the given context, whether we refer 
to the set $C$ or $P$, we omit the superscript $C$ or $P$. ${\cal I}^{C(P)}$ denotes the $6$-dimensional span of $I_{(x,y,z)(x,y,z)}^{C(P)}$, while ${\cal I}_0^{C(P)}$ denotes the $5$-dimensional  subspace of ${\cal I}^{C(P)}$ spanned by 
$\{I_{xy}^{C(P)}, I_{xz}^{C(P)}, I_{yz}^{C(P)}, I_{xx}^{C(P)}-I_{yy}^{C(P)}, 
I_{yy}^{C(P)}-I_{zz}^{C(P)}\}$. Generalizing this notation, we shall denote by $I_{x (n_x \text{times}) y (n_y \text{times}) z(n_z \text{times})}$ the sum of symmetric tensor products with $n_x$ $\sigma_x$'s, $n_y$ $\sigma_y$'s and $n_z$, $\sigma_z$'s. We omit the zeros. Therefore, for instance,  $S_x:=I_x$ and, for $n=3$, $I_{xxx}:=\sigma_x 
\otimes \sigma_x \otimes \sigma_x$.   
\bl{II0}
\be{firstAAA}
[{\cal S}, i {\cal I}]=[{\cal S}, i {\cal I}_0]=i{\cal I}_0.  
\ee
Furthermore, if $A:=iI_{zz}$ or $iI_{xx}$ or $iI_{yy}$,
%\footnote{Here we use the notation for a vector space ${\cal V},$  $ad_{\cal S}^0 {\cal %V}:={\cal V}$, $ad_{\cal S} {\cal V}:=[{\cal S}, {\cal V}]$.}
\be{second}
\left[ {\cal S}, \texttt{span} \{A\} \right] \oplus \left[ {\cal S}  ,  [{\cal S}, \texttt{span} \{A\}] \right]=i{\cal I}_0. 
\ee 
\el

\bpr
Formula (\ref{firstAAA}) follows by direct verification using the indicated bases for ${\cal I}$, ${\cal I}_0$  and ${\cal S}$. For the second property take for example $iI_{zz}$. We have  
$$
[iS_x, iI_{zz}] \vDash iI_{yz}, \qquad [iS_y, iI_{zz}]\vDash iI_{xz}, 
$$
in $\left[{\cal S}, \texttt{span} \{A\}\right] $, and 
$$
[iS_x, iI_{yz}] \vDash iI_{zz}-iI_{yy}, \, \,
[iS_y, iI_{xz}] \vDash iI_{zz}-iI_{xx},$$
$$
[iS_x, iI_{xz}] \vDash iI_{xy}, 
$$
which are in $\left[ {\cal S}\, ,  \left[{\cal S}, \texttt{span} \{A\}\right] \right]$. 
\epr 
With ${\cal L}^G$, we denote the full Lie algebra of matrices in $u(\hat n)$,  which commute with the symmetric group $S_{\hat n}$. The dimension of ${\cal L}^G$ was calculated in \cite{NoiGHZ} and it is given by ${\hat n+3} \choose {\hat n}$. With ${\cal L}$,  we shall denote the Lie algebra generated by $i\{S_x,S_y,S_z,I_{xx}-I_{yy}, I_{yy}-I_{zz}\}$. The following fact was one of the main results of \cite{NoiGHZ}. 
\bt{FromGHZ} Consider $\hat n$ spin $\frac{1}{2}$ particles and $I_{zz}$, 
$S_{(x,y,z)}$ matrices of the corresponding dimension $2^{\hat n}$. Then 
$iI_{zz}$, 
$iS_{(x,y,z)}$, generate all ${\cal L}^G\cap su(2^{\hat n})$.
\et 
In Theorem \ref{FromGHZ}, $I_{zz}$ models the {\it Ising interaction} between  {\it each} pair of spins in a network, while $S_{(x,y,z)}$ models the interaction with the external control magnetic field in the $(x,y,z)$ direction, respectively.

The matrix $J:=I_{xx}+I_{yy}+I_{zz}$, which models an {\it Heisenberg interaction} for  each pair of spins, will be important in our description of the dynamical Lie algebra for the system studied here. The Lie algebra ${\cal L}$ above defined is the same as ${\cal L}^G \cap su(2^{\hat n})$ except for $i J$. More precisely: 
\bp{Almost}
\be{ToBus1}
{\cal L}^G \cap su(2^{\hat n})={\cal L} \oplus \texttt{span} \{ i J\}. 
\ee
\ep
\bpr The inclusion $\supseteq$ follows from the fact that $iJ$ commutes with every permutation and so do the generators of ${\cal L}$, which are 
$iS_{(x,y,z)}$ and  $\{i(I_{xx}-I_{yy}), i(I_{yy}-I_{zz})\}$, and therefore all of ${\cal L}$. Moreover both $J$ and the generators of ${\cal L}$ are in $su(2^{\hat n})$. To show the inclusion $\subseteq$ it is enough to show that a set of generators of  
 ${\cal L}^G \cap su(2^{\hat n})$ belongs to ${\cal L} \oplus \texttt{span} \{ i J \}$. For this, we use Theorem \ref{FromGHZ}, and take as generators $iS_{(x,y,z)}$ and $iI_{zz}$. The matrices $iS_{(x,y,z)}$  are already in ${\cal L}$ by definition of  ${\cal L}$.
 Since 
 $$
 iI_{zz} = -\frac{1}{3} i(I_{xx}-I_{yy})-\frac{2}{3} i(I_{yy}-I_{zz})+\frac{1}{3} iJ
 $$
and   $\{ i(I_{xx}-I_{yy}), i(I_{yy}-I_{zz})\}$ are also in ${\cal L}$, we have that 
$iI_{zz}$ 
belongs to ${\cal L} \oplus \texttt{span} \{ iJ \}$. 
\epr 
\noindent We shall also use the following property of the matrix $J$. 
\bl{CommuJ}
The matrix $iJ$ commutes with ${\cal L}$. 
\el
\bpr We only need to prove that $iJ$ commutes with the generators of ${\cal L}$. We start with $iS_{(x,y,z)}$. 
By symmetry we only need to consider one among $iS_{(x,y,z)}$. Take 
$iS_x$, and calculate  $[iS_x, iJ]=[iS_x, i(I_{xx}+I_{yy}+I_{zz})]= 
[iS_x, i(I_{yy}+I_{zz})]=[iS_x, iI_{yy}]+[iS_x, iI_{zz}]$. The first term, using (\ref{commurel}) gives $iI_{zy}$ (it is clear that it contains sum of matrices with all identities except in two positions, one occupied by $\sigma_z$ and one occupied by $\sigma_y$; moreover it has to be invariant under permutations and the only matrices with this property are proportional to $iI_{yz}$; the fact that the proportionality factor is  $1$ follows from the fact that $\sigma_z$ in the first place can only occur once). { Using again (\ref{commurel}), the second term gives $- iI_{zy}$, thus these two terms sum up to zero}.

As for $i(I_{xx}-I_{yy})$ and $i(I_{yy}-I_{zz})$, again by symmetry, we need to consider only one of them. We consider  $i(I_{xx}-I_{yy})$. We have 
$[i(I_{xx}-I_{yy}), iJ]=[i(I_{xx}-I_{yy}), i(I_{xx}+I_{yy}+I_{zz})]=
[iI_{xx},iI_{yy}]+[iI_{xx},iI_{zz}]-[iI_{yy},iI_{xx}]-[iI_{yy},iI_{zz}]=$
\be{tobeus}
2[iI_{xx},iI_{yy}]+[iI_{xx},iI_{zz}]-[iI_{yy},iI_{zz}].
\ee 
In the commutator $[iI_{xx},iI_{yy}]$, writing $I_{xx}$ and $I_{yy}$ as symmetric sums of tensor products   the only terms that do not give  zero are the ones where  the two positions in $I_{xx}$ occupied by $\sigma_x$ and the two positions of $I_{yy}$ occupied by $\sigma_y$ have only one index in common (e.g., positions $(1,2)$ and position $(2,3)$). The commutator gives a term with a single $\sigma_x$, a single $\sigma_y$ and a single $\sigma_z$. Using the fact that the Lie bracket has to be permutation invariant, we obtain that $[iI_{xx},iI_{yy}]$ must be proportional to $iI_{xyz}$. The proportionality factor is in fact $1$. This can be seen by writing $I_{xx}$ as $I_{xx}=\sigma_x\otimes S_x+{\bf 1} \otimes I_{xx}^{n-1}$, where $I_{xx}^{n-1}$ is $I_{xx}$ but on $n-1$ positions, and, analogously $I_{yy}=\sigma_y\otimes S_y+{\bf 1} \otimes I_{yy}^{n-1}$. Taking the commutator one can see that the coefficients of the terms having $\sigma_z$ in the first place is $1$, and therefore, by permutation symmetry this is the coefficient if $iI_{xyz}$ as well. With an analogous reasoning, the commutator $ [iI_{xx},iI_{zz}]$ in (\ref{tobeus}) gives $-iI_{xyz}$ and the commutator $[iI_{yy},iI_{zz}]$ in (\ref{tobeus}) gives $iI_{xyz}$, so that the sum in (\ref{tobeus}) gives zero. 
 \epr
\noindent Using Proposition \ref{Almost}, we have 
\bc{commuext}
The matrix $iJ$ commutes with ${\cal L }^G$. 
\ec 
\bl{lastminute}
If $n=2$, for each $A \in {\cal L}$, 
\be{verystrong}
JA=AJ=A
\ee 
\el
\bpr
Formula (\ref{verystrong}) can be directly verified for the generators of ${\cal L}$ using (\ref{relations}), and it is extended to commutators by $J[A,B]=J(AB-BA)=(JA)B-(JB)A=AB-BA=[A,B].$ 
\epr
\subsection{The model}
We consider a network of spin $\frac{1}{2}$ particles divided into two sets, $C$ and $P$. Each spin in the set $C$ interacts via Ising interaction with each spin in the set $P$ but there is no (significant) interaction within spins in the set $C$ ($P$). The system is controlled by a common electro-magnetic field which is arbitrary in the $x$ and $y$ direction. Up to a proportionality factor, the quantum mechanical Hamiltonian of the system can be written as 
\be{BasicHam}
\begin{array}{cl}
H= &S_{z}^C \otimes S_z^P+u_x(\gamma_C S_x^C \otimes {\bf 1}+\gamma_P {\bf 1} \otimes S_x^P) +\\
& +u_y(\gamma_C S_y^C \otimes {\bf 1}+\gamma_P {\bf 1} \otimes S_y^P). 
 \end{array}
 \ee  
Here the term $S_{z}^C \otimes S_z^P$ models the Ising interaction of each spin of the set $C$ with each spin of the set $P$. This should not be confused with a term of the form $I_{zz}$ which models Ising interaction between 
{\it any pair} of spin in a network. The functions  $u_x:=u_x(t)$ and $u_y:=u_y(t)$ are control electromagnetic fields in the $x$ and $y$ directions. The parameters $\gamma_C$ and $\gamma_P$ are (proportional to) the {\it gyromagnetic ratios} of the spins in set $C$ and set $P$, respectively. The dimensions of the identity matrices ${\bf 1}$ in (\ref{BasicHam}) are $2^{n_c}$ or $2^{n_p}$, according to weather ${\bf 1}$ is on the left or on the right, respectively, of the tensor product. The Schr\"odinger equation for the system takes  the form (\ref{QSbasic}) where $A+\sum_{j}B_j u_j=-iH$ with $H$ in (\ref{BasicHam}).

\subsection{Dynamical Lie algebra and subspace controllability} \label{PIPPO}
We want to describe  the possible evolutions that can be obtained by changing the controls in (\ref{BasicHam}) and therefore we want to describe the dynamical Lie algebra ${\cal G}$ generated  by 
$$
\{ iS_{z}^C \otimes S_z^P, 
i(\gamma_C S_x^C \otimes {\bf 1}+\gamma_P {\bf 1} \otimes S_x^P), 
i(\gamma_C S_y^C \otimes {\bf 1}+\gamma_P {\bf 1} \otimes S_y^P) \}.
$$
Once ${\cal G}$ is determined,  its elements will take,  in appropriate coordinates, a block diagonal form which describes the  {\it sub-representations} of ${\cal G}$. {The Hilbert space ${\cal H}$ for the quantum state is accordingly  decomposed into  invariant  subspaces. Subspace controllability is verified if, on each subspace,  ${\cal G}$ acts as $u(m)$ or $su(m)$ where $m$ is the dimension of the given subspace.} 
Our problem is to determine the Lie algebra ${\cal G}$ and then find all its sub-representations and prove subspace controllability.

As a preliminary step,  we remark that, letting 
$$
W=[i\gamma_C S_x^C \otimes {\bf 1} + i \gamma_P {\bf 1} \otimes S_x^P, i\gamma_C S_y^C \otimes {\bf 1} + i \gamma_P {\bf 1} \otimes S_y^P],
$$
then
$$
[i\gamma_C S_x^C \otimes {\bf 1} + i \gamma_P {\bf 1} \otimes S_x^P,W ]
\vDash i\gamma_C^3 S_y^C \otimes {\bf 1}+ i \gamma_P^3 {\bf 1} \otimes S_y^{P}. 
$$ 
Therefore, since the Lie algebra contains $i\gamma_C S_y^C \otimes {\bf 1}+ i \gamma_P {\bf 1} \otimes S_y^{P}$ also, assuming $|\gamma_C| \not=  |\gamma_P|$, we have that $i S_y^C \otimes {\bf 1}$ and $i{\bf 1} \otimes S_y^{P}$ belong to ${\cal G}$. Taking the Lie brackets of  $i\gamma_C S_x^C \otimes {\bf 1} + i \gamma_P {\bf 1} \otimes S_x^P$ with $i S_y^C \otimes {\bf 1}$ and    $i{\bf 1} \otimes S_y^{P}$,   we obtain   that  $i S_z^C \otimes {\bf 1}$ and 
$i{\bf 1} \otimes S_z^{P}$ are in ${\cal G}$, and taking the Lie bracket between  $i S_y^C \otimes {\bf 1}$ ($i{\bf 1} \otimes S_y^{P}$ ) and   $i S_z^C \otimes {\bf 1}$ ($i{\bf 1} \otimes S_y^{P}$) we obtain $i S_x^C \otimes {\bf 1}$ ($i{\bf 1} \otimes S_x^{P}$ ). Therefore 
${\cal G}$ contains the $3-$dimensional subspaces 
\be{AandC}
{\cal A}^C:=\texttt{span} \{ i S_{(x,y,z)}^C \otimes {\bf 1}\}, \, \, {\cal A}^P:=\texttt{span} \{ i   {\bf 1} \otimes S_{(x,y,z)}^P\}, 
\ee
under the assumption that $|\gamma_C| \not= |\gamma_P|$. We shall assume this to be the case in the following. Therefore the dynamical Lie algebra ${\cal G}$ is the Lie algebra generated by ${\cal A}^C$, ${\cal A}^P$ and $i S_{z}^C \otimes S_z^P$.

\section{Description of the Dynamical Lie Algebra}\label{DDLA} 

\subsection{Results for general $n_c \geq 1$}

Consider the group $\hat G$, $\hat G:=S_{n_c} \otimes S_{n_p}$, where $S_{n_c}$ is the group of permutation matrices (symmetric group) on the first $n_c$ positions, corresponding to spins of the type $C$ and  
$S_{n_p}$ is the group of permutation matrices (symmetric group) on the second $n_p$ positions, corresponding to spins of the type $P$. {This means, for $C$ (and analogously for $P$) that if $Q$ is a matrix in $S_{n_c}$ and $A$ belongs to $u(2^{n_c})$ 
$QAQ^{-1}$ is obtained from $A$ by (possibly) permuting certain positions in the tensor products which appear once one expands  $A$ in the standard (tensor product type) of basis 
in $u(2^{n_c})$.}  This is a {\it group of symmetries} for the system described by the Hamiltonian 
(\ref{BasicHam})  since for every element $Q_C \otimes Q_P \in S_{n_c} \otimes S_{n_p}$, we have  
$$
[iS_{z}^C \otimes S_z^P, Q_C\otimes Q_P ]=0, 
$$
$$
[i(\gamma_C S_x^C \otimes {\bf 1}+\gamma_P {\bf 1} \otimes S_x^P),  Q_C\otimes Q_P ]=0, 
$$
$$
[i(\gamma_C S_y^C \otimes {\bf 1}+\gamma_P {\bf 1} \otimes S_y^P),  Q_C\otimes Q_P ]=0. 
$$
The generators of ${\cal G}$ all commute with $\hat G$ and therefore all of $\cal G$ commutes with $\hat G$. This implies that the dynamical Lie algebra ${\cal G}$ must be a Lie  subalgebra of the maximal subalgebra ${\cal L}^{\hat G}$ of $u(2^{n_c+n_p})$ which commutes with $\hat G$. We have 
${\cal L}^{\hat G}=i{\cal L}^{GC} \otimes {\cal L}^{GP}$. Here  ${\cal L}^{GC}$ (${\cal L}^{GP})$ is the Lie subalgebra of $u(2^{n_c})$ ($u(2^{n_p})$)  invariant under $S_{n_c}$ 
($S_{n_p}$). Therefore a basis of ${\cal L}^{\hat G}$ can be obtained by taking tensor products of a basis of ${\cal L}^{GC}$ with a basis of ${\cal L}^{GP}$ and the dimension of 
${\cal L}^{\hat G}$ is $M(n_c)M(n_p)$, where $M(n):={{n+3} \choose {n}}$ (from \cite{NoiGHZ}).  In fact, ${\cal G}$ is a Lie subalgebra of a slightly smaller Lie algebra.
\bl{nuovo}
The Lie algebra 
\be{hatL}
\hat {\cal L}= \left( i{\cal L} \otimes {\cal L}^G \right) + \left( i{\cal L}^G \otimes {\cal L} \right),
\ee 
 is a super Lie algebra of ${\cal G}$. 
 \el
 \bpr
 To see that (\ref{hatL}) is a Lie algebra, we can notice that it is the orthogonal complement in ${\cal L}^G \otimes {\cal L}^G$ to the Abelian Lie algebra 
$$
{\cal J}:=\texttt{span}\{i{\bf 1} \otimes {\bf 1}, i{\bf 1} \otimes {J}, i{J} \otimes {\bf 1}, i{J} \otimes {J}\}, 
$$
for the appropriate dimensions of the identity ${\bf 1}$ and $J$ on the left and on the right (this in the case $n_c=1$ reduces to ${\cal J}:=\texttt{span}\{i{\bf 1} \otimes {\bf 1}, i{\bf 1} \otimes {J}\}$) and commutes with it because of Lemma \ref{CommuJ} and Corollary \ref{commuext}. Therefore is closed under commutation from Lemma \ref{commut56}. Moreover all generators of ${\cal G}$ belong to ${\hat {\cal L}}$. 
\epr
We shall see that in the case $n_c=1$, ${\cal G}=\hat {\cal L}$, while for $n_c=2$ ${\cal G}\not=\hat {\cal L}$.  We now identify  certain subspaces of $\hat {\cal  L}$ which belong to  the dynamical Lie algebra ${\cal G}$.

\bp{BandD}
The following vector spaces belong to ${\cal G}$: 
\be{BandD2}
\begin{array}{lcl}
{\cal B}&:=&\texttt{span}\{ i S_{x,y,z}^C \otimes S_{x,y,z}^P \},  \\
{\cal D}_1&:=&\texttt{span} \{i S_{(x,y,z)}^C \otimes I_{(x,y,z)(x,y,z)}^P \}, \\
 {\cal D}_2&:=&\texttt{span} \{i I_{(x,y,z)(x,y,z)}^C \otimes S_{(x,y,z)}^P \}.\\
 \end{array}
\ee
\ep 
\vspace{.2cm}
\br{nuovoremark}
Notice that the above subspaces have the following dimensions: $\dim({\cal B})=9$, 
$\dim ({\cal D}_1)=18$ unless the set $P$ has cardinality $1$, in which case 
${\cal D}_1=\{0\}$, $\dim ({\cal D}_2)=18$ unless the set $C$ has cardinality $1$, in which case  
${\cal D}_2=\{0\}$. 
\er
\bpr The indicated basis of ${\cal B}$ can be obtained from $i S_{z} \otimes S_{z}$ by taking Lie brackets with elements of the basis of ${\cal A}^C$ and ${\cal A}^P$ indicated in the definition (\ref{AandC}).  Now assume that the set $P$ has cardinality strictly bigger than $1$ and take the Lie bracket of the two elements in ${\cal B}$, $i S_x \otimes S_z$ and  
$i S_y \otimes S_z$, which is $[i S_x, iS_y] \otimes S_z^2 \vDash i S_z \otimes 
({\bf 1} + I_{zz})$. Since we know that $iS_z \otimes {\bf 1}$ is in ${\cal G}$, as  it belongs to ${\cal A}^C$, we have that $iS_z \otimes I_{zz}$ is in ${\cal G}$. By taking Lie brackets with $i S_x^C \otimes {\bf 1}$ and $i S_{y}^C \otimes {\bf 1}$ we obtain 
$iS_{(x,y,z)} \otimes I_{zz}$. By taking (possibly) repeated Lie brackets with ${\bf 1} \otimes S_{(x,y,z)}^P$ (using possibly the fact that $iS_{(x,y,z)} \otimes I_{zz}$ 
belongs to ${\cal G}$) we obtain all other elements of the form $iS_{(x,y,z)} \otimes I_{(x,y,z) (x,y,z)}$. Analogously we obtain the elements in the indicated basis of ${\cal D}_2$.  
\epr

\subsection{Dynamical Lie algebra for $n_c=1$}

In the case $n_c=n_p=1$, ${\cal A}^C \oplus {\cal A}^P \oplus {\cal B}$ 
is equal to $su(4)$, so that ${\cal G}=su(4)$. In this case the system is completely controllable and our analysis terminates here. We shall therefore assume that $n_p > 1$, and 
therefore ${\cal D}_1 \not=\{0\}$ in (\ref{BandD2}) while 
${\cal D}_2 =\{0\}$.

Take $B$ in ${\cal S}^P$ and $D$ in  ${\cal I}^P$. The Lie bracket of the matrices 
$S_z \otimes  B:=\sigma_z \otimes  B  \in {\cal B}$ and 
$S_z \otimes iD=\sigma_z \otimes i D  \in {\cal D}_1$ gives 
\be{refer1}
\left[ S_z \otimes  B, S_z \otimes i D \right] =S_z^2 \otimes [B,i D] = {\bf 1}  \otimes  R, 
\ee 
for an arbitrary $R$ in $i{\cal I}_0^P$ according to (\ref{firstAAA}) of Lemma \ref{II0}. We have therefore:  
\bl{Lemma5}
If $n_c=1$, the dynamical Lie algebra  ${\cal G}$ contains 
\be{casenc1}
{\cal E}_1:={\bf 1} \otimes i{\cal I}_0^P. 
\ee 
\el 

%We are now ready to characterize the dynamical Lie algebra ${\cal G}$ in the case $n_c=1$. 
\bt{nc1}
If $n_c=1$ and for any $n_p \geq  2$ the dynamical Lie algebra ${\cal G}$ is given by 
\be{dynaLia}
{{\cal {G}}}:=\left( (\texttt{span}  \,  \{\sigma_{x,y,z}\})\otimes  {\cal L}^G  \right)  \oplus 
\left(  ({\texttt{span}} \, \{ {\bf 1} \} )
\otimes {\cal L} \right)=\hat {\cal L}. 
\ee
\et 

\bpr Using elements in ${\cal E}_1$ and elements of ${\cal A}^P$, since 
${\cal L}$ is the Lie algebra generated by $i{\cal I}_0$ and $iS_{(x,y,z)}$ we obtain anything in 
$(\texttt{span} \{ {\bf 1} \} )
\otimes {\cal L}$. Now we know from Theorem \ref{FromGHZ} that $i{I}_{zz}^P$, $iS_{(x,y,z)}^P$  and $i{\bf 1}$ generate all of ${\cal L}^{GP}$. Therefore, basis elements of ${\cal L}^{GP} \cap su(2^{ n_p})$ are obtained by (repeated) Lie brackets of $iI_{zz}^P$ and  $iS_{(x,y,z)}^P$. Define the `{\it depth}' of a basis element $K_1$ as the number of Lie brackets to be performed to obtain $K_1$. In particular, the generators $i I_{zz}$, $iS_{(x,y,z)}$ are element of depth zero. We show by induction on the depth of the basis element $K_1$ that all elements of the form 
$\sigma_{(x,y,z)} \otimes K_{1}$ can be obtained. For depth zero, we already have $i \sigma_{(x,y,z)}\otimes S_{(x,y,z)} \in {\cal B}$ and $i \sigma_{(x,y,z)} \otimes I_{zz} \in {\cal D}_1$, from Proposition \ref{BandD}. For depth $d \geq 1$, assume by induction that we have all elements $i \sigma_{(x,y,z)} \otimes K_1$ for $K_1$ in the basis of ${\cal L}^G \cap su(2^{n_p})$, $K_1$  of depth $d-1$.  If $K_2=[K_1, iS_{(x,y,z)}],$  we can obtain
$$
\begin{array}{rl}
[\sigma_{(x,y,z)} \otimes K_1, i{\bf 1} \otimes S_{x,y,z}]= &\sigma_{(x,y,z)} \otimes [K_1, iS_{(x,y,z)}]  \\
= &\sigma_{(x,y,z)} \otimes K_2.  
\end{array}
$$    
If $K_2:=[K_1, iI_{zz}]$, write  \\
$iI_{zz}=\frac{1}{3}i(I_{xx}-I_{yy}) -\frac{2}{3} i (I_{xx} - I_{zz}) +\frac{1}{3} iJ$, so that 
$$
\begin{array}{rl}
K_2= &\left[ K_1, \frac{1}{3}i(I_{xx}-I_{yy}) -\frac{2}{3} i (I_{xx} - I_{zz}) +\frac{1}{3} iJ\right]
\\
=& \left[K_1, \frac{1}{3}i(I_{xx}-I_{yy}) -\frac{2}{3} i (I_{xx} - I_{zz}) \right].
\end{array}
$$
This is true because $iJ $ commutes with ${\cal L}^G$ according to Corollary \ref{commuext}. This shows that $K_2 \in [K_1, {\cal L}]$ and since we have $\sigma_{(x,y,z)} \otimes K_1 \in {\cal G}$ (by inductive assumption) and ${\bf 1} \otimes {\cal L} \in {\cal G}$ (because we showed it above), we have 
$$
\begin{array}{rl}
\left[\sigma_{(x,y,z)} \otimes K_1, {\bf 1} \otimes i I_{zz} \right]= & \sigma_{(x,y,z)} \otimes [K_1, iI_{zz}] \\
 \in & \sigma_{(x,y,z)} \otimes [K_1, {\cal L}] \in {\cal G}. 
 \end{array}
$$
These  arguments show that, in (\ref{dynaLia}), the right hand side is included in the left hand side. We already  know that ${\cal G} \subseteq \hat {\cal L}$ by Lemma \ref{nuovo}, so the theorem is proved. 
\epr 

\subsection{Dynamical Lie algebra for $n_c=2$}
We start with some considerations for general $n_p \geq n_c=2$. Then we will give separate results for the case $n_p=2$ and $n_p>2$. 

\bl{recuperato}
If $n_c = 2$, ${\cal G}$ contains the spaces 
\be{casencm1}
i{\cal I}_0^C \otimes {\cal I}_0^P,\quad ({\bf 1}+\frac{1}{3}J) 
\otimes i {\cal I}_0^P. 
\ee
\el
\bpr
Using 
(\ref{refer1}), we have  $S_z^2 \otimes [B,iD]=({\bf 1}+I_{zz})\otimes R \in {\cal G}$, for each $R \in i{\cal I}_0^P$. Taking  (repeated) Lie brackets with elements of the form 
${\cal A}^C \subseteq {\cal G}$ and using formula (\ref{second}) of Lemma 
\ref{II0} with $A:=iI_{zz}^C$ we obtain the first one of (\ref{casencm1}). Repeating the calculation in (\ref{refer1}) with $S_z$ replaced by $S_x$ or $S_y$, we obtain $ i({\bf 1} + I_{xx}) \otimes {\cal I}_0^P \in {\cal G}$ and $ i({\bf 1} + I_{yy})\otimes {\cal I}_0^P \in {\cal G}$, which together with the corresponding  one for $x$ gives the second one in (\ref{casencm1}). 
\epr 

%%%%%%%%%%%%%%%%%%%%%%%%%%%%%%%%%
%%%%%%%%%%%%%%%%%%%%%%%%%%%%%%%%%
%%%%%%%%%%%%%%%%%%%%%%%%%%%%%%%%%

\bp{first} 
If $n_c=2$ and for all  $n_p\geq 2$, it holds that:
\be{first1}
S_{x,y,z} \otimes A,
\ee
with $A\in {\cal L}^G$ belongs to ${\cal G}$.
\ep
\bpr
The proof is by  induction on the depth of $A$, with the generators $iS_{x,y,z}$, $iI_{zz}$ and $i{\bf 1}$ of ${\cal L}^G$. 
We know that the  matrices:
\[
iS_{x,y,z}\otimes {\bf{1}} ,  \ \  \  iS_{x,y,z}\otimes S_{x,y,z}, \  \  \ iS_{x,y,z}\otimes I_{zz},
\]
are in ${\cal G}$, since the first type belongs to ${\cal A}^C$ in (\ref{AandC}), the second type belongs to ${\cal B}$ 
and the third one to ${\cal D}_1$ in (\ref{BandD2}). 
Thus equation (\ref{first1}) holds for $A$ of depth 0. Assume that it holds for all $B$ of depth $k$. If $A$ has depth $k+1$, then
either $A=[B, S_{x,y,z} ]$ or $A=[B, I_{zz}]$, and $S_{x,y,z}\otimes B\in{\cal G}$ by inductive assumption.
In the first case,  we have:
\[
[S_x\otimes B, {\bf{1}}\otimes iS_{x,y,z}] = S_x \otimes A\in {\cal G}.
\]
In the second case, we have:
\[
A=[B,iI_{zz}]=[B, \frac{1}{3}J+\frac{1}{3}i(I_{xx}-I_{yy}) -\frac{2}{3} i (I_{xx} - I_{zz})]=[B, \frac{1}{3}i(I_{xx}-I_{yy}) -\frac{2}{3} i (I_{xx} - I_{zz})],
\]
since $J$ commutes with $B$ because of Corollary \ref{commuext}. 
We also have 
\[
\left({\bf{1}} +I_{xx} \right) \otimes \frac{1}{3}i(I_{x,x}-I_{y,y}) -\frac{2}{3} i (I_{x,x} - I_{z,z})\in {\cal G},
\]
because of (\ref{casencm1}). Therefore we calculate   
\[
[ S_x\otimes B, \left({\bf{1}} +I_{xx} \right) \otimes \frac{1}{3}i(I_{x,x}-I_{y,y}) -\frac{2}{3} i (I_{x,x} - I_{z,z}) ]=
\]
\[=
1/2 \{S_x, \left({\bf{1}} +I_{xx}\right)\} \otimes [B, \frac{1}{3}i(I_{x,x}-I_{y,y}) -\frac{2}{3} i (I_{x,x} - I_{z,z}) ]= (S_x + S_xI_{xx}) \otimes [B,iI_{zz}]=
 (S_x + S_xI_{xx}) \otimes A \in {\cal G}.
\]
  Since for  $n_c=2$, $S_xI_{xx}=S_x$, we have $S_x \otimes A \in {\cal G}$, and analogously for $S_y \otimes A$ and $S_z \otimes A$. 
\epr

\bp{seconda} 
If $n_c=2$, then all matrices of the type 
\be{seconda1}
(I_{xx}-I_{zz} )\otimes A, \  \  \text{ and } \ \  (I_{yy}-I_{zz} )\otimes A
\ee
with $A\in {\cal L}$ belong to ${\cal G}$.
\ep
\bpr
We will prove the statement by induction on the depth of the matrix $A$, by taking $iS_{x,y,z}$ and $i(I_{xx}-I_{zz})$ and $i(I_{yy}-I_{zz})$ as generators of ${\cal L}$ (by definition).
By Lemma \ref{BandD}, we know that all matrices:
\[
i(I_{xx}-I_{zz} )\otimes S_{x,y,z}, \  \  \ i(I_{yy}-I_{zz} )\otimes S_{x,y,z}
\]
are in ${\cal G}$. Moreover from equation (\ref{casencm1}) we get also that the matrices:
\[
i(I_{xx}-I_{zz} )\otimes (I_{xx}-I_{zz} ),  \ \ \ i(I_{xx}-I_{zz} )\otimes (I_{yy}-I_{zz} ),
\]
and
\[
i(I_{yy}-I_{zz} )\otimes (I_{xx}-I_{zz} ),  \ \ \ i(I_{yy}-I_{zz} )\otimes (I_{yy}-I_{zz} ),
\]
are in ${\cal G}$. Thus the elements (\ref{seconda1}) are in ${\cal G}$, 
when $A$ is of depth  $0$.

On the other hand, if the depth of $A \in {\cal L}$ is $k > 0$, then either  $A=[B, i S_{x,y,z} ]$ or  $A=[B, i (I_{xx}-I_{zz})]$ 
or $A=[B, i(I_{yy}-I_{zz})]$, for $B \in {\cal L}$ of depth $k-1$. In the first case, we have:
\[
[(I_{xx}-I_{zz} )\otimes B, {\bf{1}}\otimes S_{x,y,z}] =(I_{xx}-I_{zz} ) \otimes A\in {\cal G}, 
\]
and similarly also $(I_{yy}-I_{zz} ) \otimes A\in {\cal G}$. For the second case, we know from Proposition \ref{first} that $
S_x\otimes B\in {\cal G}
$, and from Lemma \ref{BandD}, $
iS_x\otimes (I_{xx}-I_{zz} )\in {\cal G}.
$
Thus 
\[
[S_x\otimes B, S_x\otimes i(I_{xx}-I_{zz} )]= S_x^2
 \otimes [B, i (I_{xx}-I_{zz})]=  2({\bf{1}} + I_{xx})\otimes A \in {\cal G}.
\]
Using $S_z$ instead of $S_x$, we get also also the matrix $ 2( {\bf{1}} + I_{zz})\otimes A$ is in ${\cal G}$.
Thus also $(I_{xx}-I_{zz} )\otimes A$ is  in ${\cal G}$. Similarly, we prove that also $(I_{yy}-I_{zz} ) \otimes A\in {\cal G}$, as desired.
\epr
%Define the matrix $H:=\frac{1}{3}(I_{xx}+I_{yy}+I_{zz})$. 
\bp{viaskype}$
({\bf 1} +\frac{1}{3}J) \otimes {\cal L},  
$
belongs to ${\cal G}$. 
\ep
\bpr 
Using the last ones of (\ref{AandC}) and (\ref{BandD2}) we know that ${\cal G}$ contains 
$(1+\frac{1}{3}J) \otimes iS_{x,y,z}$. Using the second one of (\ref{casencm1}) we also know that ${\cal G}$ contains 
$(1+\frac{1}{3}J) \otimes {\cal I}_0$. Therefore, for every generator of ${\cal L}$, $A$, $(1+\frac{1}{3}J) \otimes A$ belongs to ${\cal G}$. Now for two elements of ${\cal L}$, $A$ and $B$, we have that 
$$
[({\bf 1}+\frac{1}{3}J) \otimes A, ({\bf 1}+\frac{1}{3}J) \otimes B]=({\bf 1}+\frac{1}{3}J)^2 \otimes [A,B]=
\frac{4}{3} ({\bf 1}+\frac{1}{3}J) \otimes [A,B], 
$$
since a direct calculation gives $({\bf 1}+\frac{1}{3}J)^2 =\frac{4}{3}({\bf 1}+\frac{1}{3}J)$. Therefore $(1+\frac{1}{3}J)\otimes A$ is in ${\cal G}$ whether $A$ is a generator of ${\cal L}$ or it is a Lie bracket of two elements of ${\cal L}$. This implies that it is in ${\cal G}$ for any $A$ in ${\cal L}$. 
\epr 
The following theorem summarizes the spaces included in ${\cal G}$ which we have identified so far for $n_c=2$. 
\bt{Summariz}
Assume $n_c=2$. Then the dynamical Lie algebra ${\cal G}$ contains the following subspaces: 
\begin{itemize}
\item[i)]
\be{LL}
i{\cal L} \otimes {\cal L} 
\ee
\item[ii)]
\be{Up3J}
({\bf 1}+\frac{1}{3}J) \otimes {\cal L}
\ee
\item[iii)]
\be{LupH}
{\cal L}\otimes \left({\bf 1} + \frac{2}{3n_p} J \right)
\ee
\item[iv)] 
${\cal A}^C$ and ${\cal A}^P$ from (\ref{AandC}). 
\end{itemize}
\et
\bpr
The subspace in (\ref{LL}) comes from (\ref{first1}) of Proposition \ref{first} and (\ref{seconda1}) of Proposition \ref{seconda} by taking Lie brackets of the elements in (\ref{seconda1}) with $iS_{x,y,z}\otimes {\bf 1}$ (which are in (\ref{first1})) to obtain the rest of ${\cal I}_0 \otimes {\cal L}$. For the subspace in (\ref{LupH}), recalling that in the case $n_c=2$, ${\cal L}={\cal S} \oplus i{\cal I}_0$, the part in (\ref{LupH}) with ${\cal S}$ on the right comes from (\ref{first1}).  The subspace $i{\cal I}_0 \otimes (1+\frac{2}{3n_p}J)$ can be obtained as follows: By induction on $n_p$, we have  
\be{Sx77}
(S_{x,y,z}^P)^2=n_p {\bf 1}+2I_{xx,yy,zz}^P. 
\ee
Take for instance $S_x$ for $n_p=n$ which we denote by $S_{x,n}$. We have 
$S_{x,n}=S_{x,n-1}\otimes {\bf 1}+{\bf 1} \otimes \sigma_x$, and 
$$
S_{x,n}^2=(S_{x,n-1}\otimes {\bf 1}+{\bf 1}\otimes \sigma_x)^2=
S_{x,n-1}^2\otimes {\bf 1}+S_{x,n-1}\otimes \sigma_x+S_{x,n-1}\otimes \sigma_x+{\bf 1}. 
$$
Using the inductive assumption on the first term we have 
$$
S_{x,n}^2=(n-1){\bf 1}+ 2 I_{x,x (n-1)}\otimes {\bf 1}+ 2 S_{x,n-1} \otimes \sigma_x+ {\bf 1}=
n{\bf 1}+2I_{x,x (n)}, 
$$ 
since we have collected in $I_{x,x (n)}$ the terms containing pairs $(\sigma_x, \sigma_x)$ in the first $n-1$ terms, which are in $I_{x,x (n-1)}\otimes {\bf 1}$, and the terms displaying $\sigma_x$ in the last factor, which are in $2 S_{x,n-1} \otimes \sigma_x$. Summing (\ref{Sx77}) for $x$, $y$ and $z$,  we obtain 
\be{comb34}
\frac{1}{3n_p}((S_x^P)^2+(S_y^P)^2+(S_z^P)^2)= {\bf 1}+\frac{2}{3n_p}J. 
\ee
Now using (\ref{first1}) and ${\cal D}_2$ in (\ref{BandD2}) with $A \in {\cal S}$ and $B \in i{\cal I}$, which are in ${\cal G}$, 
we have $[A\otimes S_x, B\otimes S_x]=[A,B]\otimes S_x^2$ and analogously for $y$ and $z$. Summing them all and using (\ref{comb34}), we have that in ${\cal G}$ we also have 
$[A,B]\otimes ({\bf 1}+\frac{2}{3n_p}J)$, and using (\ref{firstAAA}) of Lemma \ref{II0}, we obtain the space $i{\cal I}_0\otimes \left({\bf 1} + \frac{2}{3n_p} J \right)$ to complete (\ref{LupH}). 
\epr

\subsubsection{Case $n_p=2$}
\bt{Np2}
Assume $n_c=2$ and $n_p=2$. Then the dynamical Lie algebra is the direct sum of 
the subspaces (\ref{LL}), (\ref{Up3J}), (\ref{LupH}), ${\cal A}^C$, and ${\cal A}^P$, that is, of all subspaces listed in Theorem \ref{Summariz}. 
\et 
\bpr 
For $n_p=2$, the subspaces (\ref{LL}), (\ref{Up3J}), (\ref{LupH}), ${\cal A}^C$, and ${\cal A}^P$, listed in Theorem \ref{Summariz}, summarize as 
\be{Summariz2}
i{\cal L} \otimes {\cal L}, \quad ({\bf 1}+ \frac{1}{3} J) \otimes {\cal L}, \quad 
{\cal L} \otimes ({\bf 1} + \frac{1}{3}J), \quad {\bf 1} \otimes {\cal S}, \quad {\cal S} \otimes {\bf 1}. 
\ee
Since these spaces contain the generators of the dynamical Lie algebra ${\cal L}$, it is enough to prove that their  direct sum  is closed under commutation. Denote the direct sum of the first three spaces in (\ref{Summariz2}) as $\tilde {\cal L}$, so that we have to show that $\bar {\cal L}:={\tilde {\cal L}} \oplus {\cal A}^C \oplus {\cal A}^P$ is closed under commutation. It is obvious that $[{\cal A}^C,{\cal A}^C]$, 
$[{\cal A}^C,{\cal A}^P]$, $[{\cal A}^P,{\cal A}^P]$, $[\tilde {\cal L},{\cal A}^C]$, and $[\tilde {\cal L},{\cal A}^P]$ are all in $\bar {\cal L}$. Therefore, we only have to show that $[\tilde {\cal L}, \tilde {\cal L}] \subseteq \bar {\cal L}$. To this aim, it is useful to introduce the spaces   ${\cal O}_1:=({\bf 1}-J)\otimes i{\cal I}_0$, ${\cal O}_2:=i{\cal I}_0\otimes ({\bf 1}-J)$, so that  ${\cal O}_1\oplus {\cal O}_2$, is the orthogonal complement of ${\tilde {\cal L}} \oplus {\cal A}^C \oplus {\cal A}^P$ in $\hat {\cal L}$. Using Lemma \ref{lastminute} and $J^2=3{\bf 1}_4-2J$, one can verify that the first three subspaces in (\ref{Summariz2}) commute with ${\cal O}_1$ and ${\cal O}_2$. Therefore, the commutator of any two elements, according to Lemma \ref{commut56} is orthogonal to ${\cal O}_1$ and ${\cal O}_2$, and therefore it belongs to $\bar {\cal L}$. 
\epr 
Notice that in this case ${\cal G}$ is a {\it proper} subalgebra of $\hat {\cal L}$. 
\subsubsection{Case $n_p >2$}
\bt{Npmorethan2}
Assume $n_c=2$ and $n_p > 2$ then 
\be{Gtot}
{\cal G}=\left( i{\cal L} \otimes {\cal L}^G \right) \oplus 
\left( \left({\bf 1} + \frac{1}{3} J \right) \otimes {\cal L} \right) \oplus  \left( {\bf 1} \oplus {\cal S} \right)
\ee
\et 
\bpr 
First, we see that the right hand side is included in ${\cal G}$. The last two terms of the direct sum are in ${\cal A}^P$ in (\ref{AandC}) and in 
(\ref{Up3J}). Moreover consider $A$, an arbitrary element of ${\cal S}$, and $B$, an arbitrary element of $i{\cal I}_0$. Then $A \otimes S_x \in i{\cal L} \otimes {\cal L} \subseteq {\cal G}$ because of (\ref{LL})  and $B \otimes I_{xxx} \in 
i{\cal L} \otimes {\cal L} \subseteq {\cal G}$ because of (\ref{LL}). We calculate 
$$
[A \otimes S_x, B \otimes I_{xxx}]=[A,B] \otimes S_xI_{xxx}. 
$$
$[A,B]$ can be an arbitrary element of $i{\cal I}_0$ according to (\ref{firstAAA}) of Lemma \ref{II0}, while $S_xI_{xxx}$ is a linear combination with nonzero coefficients of $I_{xx}$ and (if $n_p \geq 4$) $I_{xxxx}$. Since $iI_{xxxx} \in {\cal L}$, $[A,B] \otimes I_{xxxx} \in i{\cal L} \otimes {\cal L}$ which is already in ${\cal G}$ because of (\ref{LL}). Therefore $[A,B] \otimes I_{xx} \in {\cal G}$. Repeating this calculation with $x$ replaced by $y$ or $z$ and summing  all the terms, we obtain that $i{\cal I}_0 \otimes J \in {\cal G}$. We also have $i{\cal I}_0 \otimes {\bf 1}$ because of (\ref{LupH}), ${\cal S} \otimes {\bf 1}$ because of (\ref{AandC}), ${\cal S} \otimes J$ because of   (\ref{LupH}) and $i{\cal L} \otimes {\cal L}$ because of (\ref{LL}).  These together give $i {\cal L} \otimes {\cal L}^G$.

To show the fact that ${\cal G}$ is included in the right hand side we  notice that all the generators of ${\cal G}$ are in the right hand side of (\ref{Gtot}). Moreover we can check the commutations of the subspaces in (\ref{Gtot}).  We report only the checks  that are not immediate. We have 
$$
[i{\cal L} \otimes {\cal L}^G, i{\cal L} \otimes {\cal L}^G]= 
[{\cal L}, {\cal L}] \otimes \{ {\cal L}^G,{\cal L}^G\} +
\{ {\cal L},{\cal L}\}\otimes [{\cal L}^G,{\cal L}^G]\subseteq 
i{\cal L} \otimes {\cal L}^G+\{ {\cal L},{\cal L}\} \otimes {\cal L}. 
$$
In the last term in the right hand side $\{ {\cal L},{\cal L}\}$ must be a linear combination of ${\bf 1}+\frac{1}{3}J$ and elements in $i{\cal L}$ becuase it is in $i{\cal L}^G$ and orthogonal to ${\bf 1}-J$ because of Lemma \ref{lastminute}. In fact, for $A$ and $B$ in ${\cal L}$, we have 
$Tr(({\bf 1}-J)(AB+BA))=Tr(AB+BA-AB-BA)=0$. Therefore these commutators 
are in the right hand side of (\ref{Gtot}). 
$$
[i{\cal L} \otimes {\cal L}^G, ({\bf 1} +\frac{1}{3}J)\otimes {\cal L}]=
\{ {\cal L}, ({\bf 1}+\frac{1}{3}J)\} \otimes [{\cal L}^G, {\cal L}]+
[{\cal L}, ({\bf 1} +\frac{1}{3}J)] \otimes \{{\cal L}^G, {\cal L} \}. 
$$
The last term is zero because of Lemma \ref{CommuJ} while the first term is in $i{\cal L} \otimes {\cal L}$ because of Lemma \ref{lastminute}. 
Moreover 
$$
\left[ ({\bf 1}+\frac{1}{3}J) \otimes {\cal L}, ({\bf 1}+\frac{1}{3}J) \otimes {\cal L} \right]\subseteq ({\bf 1}+\frac{1}{3}J)^2 \otimes {\cal L}=
\frac{4}{3} ({\bf 1}+\frac{1}{3}J) \otimes {\cal L}. 
$$
\epr 
We remark that ${\cal G}$ in (\ref{Gtot}) is always a {\it proper} 
subalgebra of $\hat {\cal L}$ in (\ref{hatL}). In fact, if ${\cal C}$ is a subspace in ${\cal L}^P$ orthogonal  to ${\cal S}$, the  subspace in $\hat {\cal L}$, $({\bf 1}-J) \otimes {\cal C}$ belongs to $\hat {\cal L}$ but it is orthogonal to ${\cal G}$ in (\ref{Gtot}). Nevertheless, we will see in the next section that subspace controllability is verified in all cases considered in this paper.

\section{Subspace Controllability}\label{SC3}

In general, if a system of the form (\ref{QSbasic}) admits a discrete 
group of symmetries $\hat G$, i.e., a group $\hat G$ such that $[A,P]=0$, $[B_j,P]=0$, $\forall P \in \hat G$, the maximal Lie subalgebra of $u(\hat n)$ which commutes with 
$\hat G$, ${\cal L}^{\hat G}$, acts on certain invariant subspaces ${\cal H}_j$ of the Hilbert space 
${\cal H}$ as $u(\dim({\cal H}_j))$. Each of such subspaces is an irreducible representation of ${\cal L}^{\hat G}$ (cf., \cite{Jonas} Theorem 4). In an appropriate basis of ${\cal H}$, therefore, ${\cal L}^{\hat G}$ can  be written in block diagonal form, where each block can take values in $u(\dim({\cal H}_j))$. The dynamical Le algebra associated with a system having $\hat G$ as a group of symmetries also displays a block diagonal form in the same basis although not necessarily equal to the full 
${\cal L}^{\hat G}$. In the preferred basis however one can study the action of the dynamical Lie algebra on each subspace and determine subspace controllability. This is the plan  
we follow here. 

A method to find the desired   basis was described in \cite{Jonas} and it uses the so-called {\it Generalized Young Symmetrizers (GYS)} where the word `Generalized' refers to the fact that, in the case where the group $\hat G$ is the symmetry group,  they reduce to the classical Young symmetrizers of group representation theory as described for instance in \cite{Tung}. More precisely, consider the representation of $\hat G$ on ${\cal H}$ and the {\it group algebra} of $\hat G$ (i.e., the algebra over the complex field generated by a basis of $\hat G$), $C[\hat G]$. Then the GYS are elements of $C[\hat G]$, and operators on ${\cal H}$, $\Pi_j$ satisfying 
C) ({\it Completeness}): $\sum_j\Pi_j={\bf 1}$; O) ({\it Orthogonality}): $\Pi_j \Pi_k=\delta_{j,k} \Pi_j$, where $\delta_{j,k}$ is the Kronecker delta; P) ({\it Primitivity}): $\Pi_j g \Pi_j=\lambda_g \Pi_j$, where $\lambda_g$ is a scalar which depends only on $g$ (and not on $j$) H) ({\it Hermiticity}): For every $j$, $\Pi_j^\dagger=\Pi_j$. If the GYS are known for a given group $\hat G$ on a Hilbert space ${\cal H}$, then the images of the various $\Pi_j \,: \, {\cal H} \rightarrow {\cal H} $ give the subspace decomposition of ${\cal H}$ which block diagonalizes the Lie algebra ${\cal L}^{\hat G}$. In the cases where $\hat G$ is the symmetric group $S_{\hat n}$ over $\hat n$ objects, the (generalized) Young symmetrizers can be found using the classical method of Young tableaux (see, e.g., \cite{Tung}) modified in references \cite{due} \cite{sedici} to meet the Orthogonality and Hermiticity requirements. A method is given in \cite{Jonas} to compute the GYS in the case where $\hat G$ is Abelian. However,  the calculation of GYS for general discrete groups is an open problem. We observe that if ${\cal H}:{\cal H}_C \otimes {\cal H}_P$ the tensor product of two Hilbert spaces ${\cal H}_C$, ${\cal H}_P$, as in bipartite quantum systems, and $\hat G$ is the product of two groups 
$\hat G:=\hat G_C \otimes \hat G_P$, with $\hat G_{C(P)}$ acting on ${\cal H}_{C(P)}$, then the GYS can be found as tensor products of GYS on ${\cal H}_{C(P)}$ for $\hat G_{C(P)}$, 
$\Pi_j^C \otimes \Pi_k^P$. It is indeed readily  verified  that if $\{\Pi_j^C\}$ and 
$ \{\Pi_k^P\}$ satisfy the requirements (C,O,P,H) above on ${\cal H}_C$ and ${\cal H}_P$, respectively, then $\{\Pi_j^C \otimes \Pi_k^P\}$ satisfy the same requirements (C,O,P,H) on ${\cal H}_C \otimes {\cal H}_P$.  The invariant subspaces are ${\cal H}_{j,k}:=(Im \, {\Pi_j^C}) \otimes (Im \, {\Pi_k^C})$ and, in this basis, the (maximal) invariant Lie algebra 
${\cal L}^{\hat G_C} \otimes   {\cal L}^{\hat G_P}$ takes a block diagonal form.

For the systems treated in this paper, the symmetry groups $\hat G_C$ and $\hat G_P$ are 
the symmetric group on $n_c$ and $n_p$ objects, respectively. The decomposition is obtained using the GYS of \cite{due}, \cite{sedici}, \cite{Tung}. 
Let $G$ be now the symmetric group and consider the matrix $J$ defined in Lemma \ref{CommuJ} and Corollary \ref{commuext} in the basis determined by the GYS. In this basis,  the elements of ${\cal L}^G$ are block diagonal and every block is an {\it arbitrary matrix}  in $u(m)$ for appropriate $m$ (cf. Theorem 2 in \cite{Jonas}).
%\footnote{Formulas are available to determine the dimension of each block (cf., e.g., formula (28) in \cite{Jonas}).} 
Since each block of the matrices in ${\cal L}^G$ can be an  arbitrary skew-Hermitian matrix of appropriate dimensions, $iJ$ is also a block diagonal matrix, i.e., 
$$
iJ:= \left[\begin{matrix} 
   iJ_1 & & \\ & \ddots & \\ & & iJ_d 
         \end{matrix}\right], 
$$
with $iJ_k$, $k=1,...,d$ commuting with the corresponding block of the matrices in ${\cal L}^G$. Since such a block defines an {\it irreducible representation} of $u(m_k)$ for appropriate dimensions $m_k$, it follows from Sch\"ur's Lemma 
(see, e.g., \cite{Fulton}) that all $iJ_k$ are scalar matrices. Consider now the matrices in ${\cal L}$ and ${\cal L}^G$ and their restrictions 
to one of the subspaces $Im \Pi_k$, of dimensions $m_k$. A basis for ${\cal L}^G$ restricted to $Im \Pi_k$  is given by a basis of $u(m_k)$ while a basis of ${\cal L}$ contains at least a basis of $su(m_k)$ since the restriction of ${\cal L}$ to  $Im \Pi_k$ differs by $u(m_k)$ at most by multiples of the identity. This is due to  Proposition \ref{Almost}, along with the fact, seen above, that $iJ$ acts as a scalar matrix on $Im \Pi_k$. 

We are now ready to conclude  subspace controllability for all the situations treated in this paper. Consider first the {\bf case $n_c=1$, and $n_p \geq 1$}, for which we have proved in Theorem \ref{nc1} that the dynamical Lie algebra is $\hat {\cal L}$ in (\ref{hatL}).   The GYS on ${\cal H}^C$ are the trivial identity, and all the invariant subspaces are ${\cal H}^C \otimes \Pi_k {\cal H}^P$, where $\Pi_k$ are the GYS's for the system $P$. A basis of ${\cal G}=\hat {\cal L}$ is given by $\sigma_{x,y,z} \otimes {\cal B}^L$, $\sigma_{x,y,z} \otimes \{i{\bf 1},i{J} \}$, ${\bf 1} \otimes {\cal B}^L$, where by ${\cal B}^L$ we have denoted a basis of ${\cal L}$. Since, as we have seen above, ${\cal L}$ acts on $\Pi_k {\cal H}^C$ as $u(m_k)$, $m_k:=\dim(\Pi_k {\cal H}^C)$, except 
possibly for multiples of the identity, a basis for the restriction of ${\cal G}$ to 
${\cal H}^C \otimes \Pi_k {\cal H}^P$, contains $\sigma_{x,y,z} \otimes {\cal U}_k$, 
 $\sigma_{x,y,z} \otimes {\bf 1}$ and ${\bf 1} \otimes {\cal U}_k$, where ${\cal U}_k$ is a basis of $su(m_k)$. Therefore it contains a basis of $su(2m_k)$ and therefore controllability is verified. Consider now the {\bf case $n_c=2$, $n_p=2$}, where the dynamical Lie algebra is described by Theorem \ref{Np2}. If ${\cal B}^L$ is a basis of ${\cal L}$, as above, a basis for ${\cal G}$ is given by $i{\cal B}^L \otimes {\cal B}^L$, $({\bf 1}+\frac{1}{3}J) \otimes {\cal B}^L$, 
 ${\cal B}^L \otimes ({\bf 1}+\frac{1}{3}J)$, 
 ${\bf 1} \otimes i \sigma_{x,y,z}$, $i\sigma_{x,y,z} \otimes {\bf 1}$. Consider two GYS $\Pi_j^C$ and $\Pi_k^P$ and the invariant space $\Pi_j^C {\cal H}^C\otimes  \Pi_j^P {\cal H}^P$ with dimensions $m_j=\dim (\Pi_j^C {\cal H}^C)$, 
 $m_k=\dim (\Pi_k^C {\cal H}^k)$. A basis for the restriction of ${\cal G}$ to 
 $\Pi_j^C {\cal H}^C \otimes  \Pi_j^P {\cal H}^P$ contains $i{\cal U}_j\otimes {\cal U}_k$, ${\bf 1} \otimes {\cal U}_k$, ${\cal U}_j \otimes {\bf 1}$, and therefore it contains a basis of $su(m_jm_k)$. Analogously, consider the {\bf case $n_c=2$, $n_p > 2$}. 
 A basis for the dynamical Lie algebra ${\cal G}$ described in Theorem 
 \ref{Npmorethan2} is, with the above notation, $i{\cal B}^L \otimes {\cal B}^L$, ${\cal B}^L \otimes \{ 1,J\}$, 
 $({\bf 1}+ \frac{1}{3} J) \otimes {\cal B}^L$, ${\bf 1} \otimes i S_{x,y,z}$ whose restriction to $\Pi_j^C {\cal H}^C \otimes Pi_j^P {\cal H}^P$ contains $i{\cal U}_j \otimes {\cal U}_k$, ${\cal U}_j \otimes {\bf 1}$, ${\bf 1} \otimes {\cal U}_k$, and therefore $su(m_jm_k)$. We have therefore with the following theorem.

\bt{nc2}
The system (\ref{QSbasic}) with one or two central spins ($n_c=1$ or $n_c=2$) with any number 
$n_p \geq n_c$ of surrounding spins, simultaneously controlled, is subspace controllable. 
\et 

\bex{Example} 
To illustrate some of the concepts and procedures described above, we consider the system of one central spin $n_c=1$ along with $n_p=3$ surrounding spins. The symmetric group on the central spin is trivial being made up of just the identity. There is a single GYS given by the identity. For the symmetric group $S_3$ on the $P$ part of the space, we obtain the GYS using the method of \cite{due}, \cite{sedici}, \cite{Tung}, 
 based on the Young tableaux. We refer to these references for details on the method. For $n=3$ there are three possible {\it partitions} of $n$  and therefore three possible {\it Young diagram} (also called {\it Young shapes}). Recall that a partition of an integer $n$ is a sequence of positive integers $\lambda_1 \geq \lambda_2 \geq \cdots \geq \lambda_d$, with $\lambda_1+\lambda_2+\cdots +\lambda_d=n$ and the corresponding Young diagram is made up of boxes arranged in rows of length 
$\lambda_1$, $\lambda_2$,...,$\lambda_d$. Therefore for $n=3$, we have the partitions $(3)$, $(2,1)$, $(1,1,1)$ which correspond to the Young diagrams 
\be{diagrammi}
\yng(3) \, , \qquad \qquad \yng(2,1)\, , \qquad \qquad \yng(1,1,1) \, , 
\ee
respectively. To each Young diagram, there  corresponds a certain number of {\it Standard Young Tableaux} obtained by filling the boxes of the Young diagram with the numbers $1$ through $n$  (3 in this case) so that they appear in strictly increasing order in the rows and in the columns. The following are the possible standard Young tableaux corresponding  to the Young diagrams in  (\ref{diagrammi}). In particular, the first one corresponds to the first diagram in (\ref{diagrammi}), the second and third  correspond to the second one in (\ref{diagrammi}) and the fourth one corresponds to the third one in (\ref{diagrammi}) 
\be{tableaux2}
\young(123) \, , \qquad \young(12,3) \, ,   \qquad  \young(13,2) \, , \qquad  \young(1,2,3) \,.  
\ee
To each tableaux there corresponds a GYS whose image is an invariant subspace for the Lie algebra representation. We refer to \cite{Jonas} for a summary of the procedure to obtain such GYS's. In our case the GYS corresponding to the first diagram in (\ref{tableaux2}) has $4-$dimensional image, the ones corresponding to the second and third have two-dimensional images and the one corresponding to the last one has zero dimensional image. Therefore the invariant subspaces for the system with one
 central spin and $n_p=3$ surrounding spin, simultaneously controlled, have dimensions $2 \times 4$, $2 \times 2$ and $2 \times 2$. 
\eex

We conclude the section by discussing the dimension of the invariant (controllable)  subspaces and how it increases with $n_p$. We recall (see, e.g., \cite{Jonas}) that there is an explicit general formula to obtain the dimension of the image of a GYS, $\Pi_T$, corresponding to a 
Young tableaux $T$. Such formula specializes to our case (where the dimension of the underlying subspace is $2$) as 
\be{formuladime}
\dim (\texttt{Im} P_T)=\frac{\prod_{l=1}^r \prod_{k=1}^{\lambda_l} (2-l+k)}{\texttt{Hook}(T)}.  
\ee 
Here $r$ is the number of rows in the Young diagram associated with $T$, 
$\lambda_l$ is the number of boxes in the $l$-th row, and  
$\texttt{Hook}(T)$ is the {\it Hook length} of the Young 
diagram associated 
with $T$. It is calculated by considering, for each box, the number of boxes directly to the right + the number of boxes directly below + 1 and then taking the product of all the numbers obtained this way. Using formula (\ref{formuladime}) it is possible to derive, for each $n_p$,  the dimensions of all invariant subspaces.
Fix $n=n_p$. From formula (\ref{formuladime}), Young diagrams with more than two rows give zero dimensional spaces. So we have to consider only Young diagrams with one or two rows. There is only one diagram with one row, $T_1$,  i.e., the diagram containing $n_p$ boxes, and in (\ref{formuladime}) $r=1$ and $\lambda_1=n$. For this diagram,  the Hook length is $n!$.  We thus have:
\[
\dim (\texttt{Im} P_{T_1})=\frac{ \prod_{k=1}^{n} (1+k)}{n!}=n+1. 
\]
For diagrams with two rows, the possible partitions are of the type $\lambda_1=n-k$ and $\lambda_2=k$, with $k$  integer and  $k\leq\frac{n}{2}$. For example 
\[
 \yng(7,3)
\]
is  the Young diagram  for the case $n=10$ and $k=3$. For the diagram corresponding to a given $k$, $T_2^k$, the Hook length is  
$${\texttt{Hook}(T_2^k)} =(n+1-k)(n-k)\cdots(n-2k+2) \cdot(n-2k)!\cdot k!.$$ Thus we have
\[
\dim (\texttt{Im} P_{T_2^k})=\frac{ \prod_{j=1}^{n-k} (1+j) \prod_{j=1}^{k} j}{(n+1-k)(n-k)\cdots(n-2k+2) \cdot(n-2k)!\cdot k!}=n-2k+1. 
\]

So, for this central spin model, the dimension of the invariant subspaces grows linearly  with $n$. The largest space has dimension $n+1$. The dimensions of the full invariant supspaces of the model with $1$ and $2$ central spins are obtained by multiplying by the dimensions obtained for ${\cal H}^P$ by the dimensions of the invariant subspaces of ${\cal H}^C$, which, with the same method of Young tableaux, can be shown to be  $2$ in the case $n_c=1$ and $1$ or $3$ in the case $n_c=2$. The largest possible dimension is therefore obtained for $n_c=2$ and it is { $3(n_p+1)$.}
This  behavior is different from the one of the system considered  in the paper \cite{ConDaniel}, where the dimension of one of the invariant subspaces grows  exponentially with the number of spins. This is essentially due to a much larger number of symmetries in our case.

%Applying this formula, we obtain that the subspace corresponding to the fourth tableaux in (\ref{tableaux2}) is zero dimensional. 
%On the other hand the subspace corresponding to the first tableaux has dimension $4$ while the two subspaces corresponding to the second and third tableaux have dimension $2$. By tensoring with the two dimensional space corresponding to the central 
%spin $C$ we obtain invariant subspaces of dimensions $8$, $4$ and $4$. From the above analysis the system is controllable on each of these subspaces and therefore subspace controllable (cf. Theorem \ref{nc2}).  

\section{Conclusions}\label{conclu}

The calculation of the dynamical Lie algebra of a quantum system is the method of choice to study its controllability properties \cite{Mikobook}. However such direct calculation might be difficult in cases of very large systems and in particular networks of spins where the dimension of the underlying full Hilbert space  grows exponentially with the number of particles. For this reason, it is important to device methods to assess controllability from the topology of the network and its possible symmetries. Symmetries, in particular, prevent full controllability and determine a number of invariant susbspaces on which the system evolves. Such invariant subspaces are obtained as images of Generalized Young Symmetrizers. Full controllability on each of these subspaces is then possible. 

In this paper we have taken the first steps in understanding such dynamical decomposition and subspace controllability for {\it multipartite} systems where different symmetry groups act on different subsystem. Motivated by common experimental situations with N-V centers in diamonds, we have considered a configuration of one or two  central spin surrounded by a number of spins. The full symmetric group acts on the central spins alone and-or on the surrounding spins alone without modifying the Hamiltonian which describes the dynamics. A common electromagnetic field is used for control. We have computed the dynamical Lie algebra and proved that such a system is subspace controllable, that is full controllability is  verified on each invariant subsystem. Quantum evolution is a parallel of the evolution of various subsystems and we can use one of them to perform various tasks of, for instance, quantum computation and-or simulation. 
%The evolution on the other sub-systems can be used, for example,  to validate the nominal evolution on the main system  in the presence of noise and imperfections in the implementation.

\vs 

\vs

\noindent {\bf Acknowledgement} D. D'Alessandro research 
is supported by NSF under Grant EECS-17890998. Preliminary results presenting only the case $n_c=1$ were submitted to the European Control Conference 2019.


\begin{thebibliography}{99}

\bibitem{NoiGHZ}  F. Albertini and D. D'Alessandro, Controllability of symmetric spin networks, 
{\it  J. Math. Phys.}  59, 052102 (2018).

\bibitem{due} J. Alcock-Zeilinger and H. Weigert, Compact Hermitian Young projection operators, {\it J. Math. Phys.}, 58(5), October 2016.  

\bibitem{Xinhua} J. Chen, H. Zhou, C. Duan, and X. Peng, Preparing GHZ and W states on a long-range Ising spin model by global control, {\it Physical Review A} (2017)

\bibitem{Mikobook}  D. D'Alessandro, {\it Introduction to Quantum Control
 and Dynamics}, CRC Press, Boca Raton FL, August 2007.

\bibitem{Jonas} D. D'Alessandro and J. Hartwig, Generalized Young Symmetrizers for the Analysis of  Control Systems on Tensor Spaces, arXiv:1806.01179.

\bibitem{Slava1} G. de Lange, T. van der Sar, M.S. Blok, Z. H. Wang, V. V. Dobrovitski and R. Hanson, Controlling the quantum dynamics of a mesoscopic spin bath in  diamond, {\it Scientific Reports 2}, 382 (2012). 

\bibitem{W} W. D\"ur, G. Vidal and J. I. Cirac, 
{\it Phys. Rev. A}, {\bf 62}, 062314 (2000)
 
\bibitem{Fulton} W. Fulton and J. Harris, {\it Representation Theory; 
A First Course}, Graduate Texts in Mathematics, No. 129, Springer, New York 2004.  
 
\bibitem{GHZ} D. M. Greenberger, M. A. Horne and A. Zeilinger, Bell's theorem, quantum theory and the conceptions of the universe, pp. 73-76, Kluwer Academics, Dordrecht, The Netherlands, (1989).  
 
\bibitem{JS} V. Jurdjevi\'c and H. Sussmann, Control systems on Lie
groups, {\it Journal of Differential Equations}, 12,  313-329,
(1972).

\bibitem{sedici} S. Keppeler and M. Sj\"odal, Hermitian Young operators, {\it Journal of Mathematical Physics}, 55, (2014) 021702. 

\bibitem{Lloyd} S. Lloyd, Almost any quantum logic gate is universal, 
{\it Physical Review Letters}, Volume 75, Number 2, July 1995.

\bibitem{NC} M. A. Nielsen and I. L. Chuang, {\it Quantum
Computation and Quantum Information}, Cambridge University Press,,
Cambridge, U.K., New York, 2000.


\bibitem{Salapaka} V. Ramakrishna, M. Salapaka, M. Dahleh, H. Rabitz, A. Peirce, Controllability of molecular systems, {\it Physical Review A}, Vol. 51, No. 2, February 1995, 960-966.

\bibitem{Slava2} T. H. Taminiau, J. Cramer, T. van der Sar, V. V. Dobrovitski, and R. Hanson, Universal control and error correction in multi-qubit spin registers in diamond, {\it Nature Nanotech.} 9, 171 (2014)


\bibitem{Tung} W. K. Tung, {\it Group Theory in Physics}, World Scientific, Singapore, 1985. 

\bibitem{ConDaniel} X. Wang, D. Burgarth, and S. Schirmer, Subspace controllability of spin $\frac{1}{2}$ chains with symmetries, {\it Physical Review A}, {\bf 94}, 052319, (2016). 

\bibitem{SenzaDaniel} X. Wang, P. Pemberton-Ross, and S. Schirmer, Symmetry and Controllability for spin networks with a single-node control, 
{\it IEEE Transactions on Automatic Control}, {\bf 57}, 1945, (2012). 

\end{thebibliography}
\end{document}